 \documentclass[smallabstract,smallcaptions]{dccpaper}

\usepackage{epsfig}
\usepackage{amsmath}
\usepackage{amssymb}
\usepackage{color}
\usepackage{url}
\usepackage{subcaption}
\newlength{\figurewidth}
\newlength{\smallfigurewidth}
\usepackage{multirow}

\begin{document}
\title
{\large
\textbf{Towards improved lossy image compression:\\Human image reconstruction with public-domain images}
}

\author{%
Ashutosh Bhown$^1$, Soham Mukherjee$^2$, Sean Yang$^3$,\\
Irena Fischer-Hwang$^4$, Shubham Chandak$^4$, Kedar Tatwawadi$^4$, \\
Judith Fan$^5$, Tsachy Weissman$^4$\\[0.5em]
{\small\begin{minipage}{\linewidth}\begin{center}
\begin{tabular}{ccc}
$^1$Palo Alto High School & \hspace*{0.5in} & $^2$Monta Vista High School \\
%Street Address One && Street Address Two \\
%City, State, ZIP, Country && City, State, ZIP, Country\\
%\url{email@address} && \url{email@address}
\end{tabular}
\end{center}\end{minipage}}\\
{\small\begin{minipage}{\linewidth}\begin{center}
\begin{tabular}{ccc}
$^3$Saint Francis High School & \hspace*{0.5in} & $^4$Stanford University \\
%Street Address One && Street Address Two \\
%City, State, ZIP, Country && City, State, ZIP, Country\\
% && \url{schandak@stanford.edu}
\end{tabular}\\
\centering $^5$ University of California San Diego
\end{center}\end{minipage}}\\
\url{schandak@stanford.edu}
}

\maketitle
\thispagestyle{empty}
\let\thefootnote\relax\footnote{Most of this work was performed as part of the first three authors' summer internship at Stanford Electrical Engineering department.}
\begin{abstract}
Lossy image compression has been studied extensively in the context of typical loss functions such as RMSE, MS-SSIM, etc. However, compression at low bitrates generally produces unsatisfying results. Furthermore, the availability of massive public image datasets appears to have hardly been exploited in image compression. Here, we present a paradigm for eliciting human image reconstruction in order to perform lossy image compression. In this paradigm, one human describes images to a second human, whose task is to reconstruct the target image using publicly available images and text instructions. The resulting reconstructions are then evaluated by human raters on the Amazon Mechanical Turk platform and compared to reconstructions obtained using state-of-the-art compressor WebP. Our results suggest that prioritizing semantic visual elements may be key to achieving significant improvements in image compression, and that our paradigm can be used to develop a more human-centric loss function.\\
\textbf{Data:} The images, results and additional data are available at \url{https://compression.stanford.edu/human-compression}.
\end{abstract}

\Section{Introduction}

\begin{figure}[!htpb]
\centering
\includegraphics[width=6in]{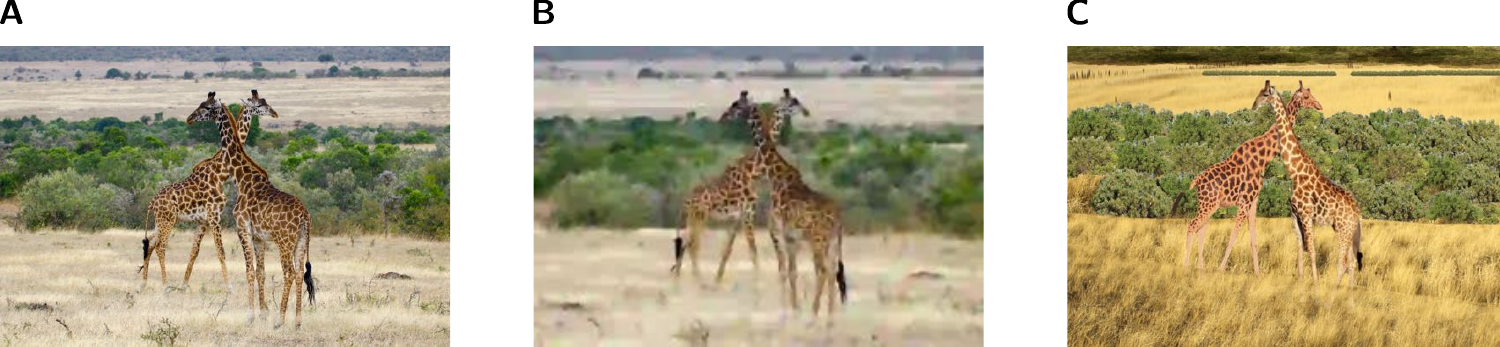}
\caption{\label{figure:giraffe_intro} Original giraffe image with WebP and human reconstructions. (A) Original giraffe image, (B) reconstruction of a WebP compressed version that is over 1,000$\times$ smaller than the original, and (C) a human reconstruction whose compressed representation is similar in size to the WebP compressed version.}
\end{figure}

Image compression is critical for achieving efficient storage and communication of image data. Today's most popular image formats and compression techniques include PNG \cite{png}, JPEG \cite{jpeg}, JPEG2000 \cite{jpeg2000}, JPEG XR \cite{jpeg_xr}, BPG \cite{bpg} and WebP \cite{WebP}. In order to achieve significant reduction in image size, most compression techniques allow some loss while compressing images. However, traditional loss functions prioritize pixel-by-pixel fidelity, leading to blurry and unnatural images at high loss levels. The left two panels of Figure \ref{figure:giraffe_intro} show an example in which compression and reconstruction using WebP \cite{WebP} result in a severely blurred image.

It is natural to posit that better compression results might be achieved by optimizing for visual properties that a human viewer cares about preserving, rather than local, pixel-level differences. The rightmost panel of Figure \ref{figure:giraffe_intro} shows an example of a reconstruction which prioritizes image content over pixel-level fidelity. Indeed, there has been a large body of work in the computer vision community~\cite{wang2004image}\cite{wang2003multiscale}\cite{chinen2018towards} towards developing loss functions that more accurately reflect human visual priorities. Some compression methods, for example, take advantage of the fact that human visual perception is more susceptible to differences in intensity than in color, and quantize color space more crudely than intensity space in order to achieve better compression performance.

Here we propose a novel strategy for improving lossy image compression that leverages human image reconstruction behavior in order to expose which visual properties humans care about preserving in images. We introduce a novel experimental paradigm for accomplishing ``human image compression'' involving two human participants who interact with one another in real time. During each interaction, the ``describer'' privately views a target image and provides natural language descriptions of the image with the goal of helping the ``reconstructor" (who cannot view the target image) produce a reconstruction of that image. Both participants have full access to publicly available images on the internet, allowing them to augment their natural language descriptions with links to images, which, in turn, may be combined during target image reconstruction. Since this experimental setup results in a set of instructions for creating a reconstruction of some target image under some definition of loss, this experimental paradigm may be thought of as accomplishing lossy compression.
 
To determine the quality of the reconstruction, we solicited human judgments about the reconstructed image using the Amazon Mechanical Turk (MTurk) platform \cite{mturk}. We used the MTurk platform to perform an evaluation experiment which assessed the quality of human-reconstructed images, and benchmarked the performance of human reconstructions against state-of-the art compressor WebP.
%The compressed size of the text chat in our framework represents the size of the compressed image, and the MTurk score is considered the (negative) ``loss" associated with human compression. 
We present the results of human compression for 13 high-resolution images of different types.
%The results show that our human compression scheme performs better than the WebP compressor on 10 out of 13 images.

\Section{Related Works}
There has been significant work on incorporating aspects of the human visual system towards improving lossy image compressors. Many commonly used compressors such as JPEG, JPEG2000 and WebP already attempt to implicitly capture properties of human perception. For example, the human visual system is prone to disregarding sharp edges in images, so JPEG quantizes high frequency components heavily. The MS-SSIM metric was developed to emulate the higher-level image similarity that seems to be valued by humans, and is used by \cite{richter2009ms} and \cite{balle2016end} for optimizing image compression. The compressor Guetzli~\cite{ginesu2012objective} includes a perceptual JPEG encoder optimized for a new image similarity metric dubbed ``butteraugli"\cite{butteraugli}. More recently, \cite{perceptual} trained a neural network to predict human perceptual quality scores on a large dataset of human-scored images. 

Another interesting line of work attempts to capture the effects of human perception by using generative models for lossy compression, which implicitly capture distributions of natural images. Then, discriminator models are used to train the generative models instead of image similarity metrics like RMSE or MS-SSIM. Furthermore, the discriminator models are themselves trained to distinguish between natural and synthetically generated images. For example, \cite{gan} uses generative adversarial networks to obtain visually pleasing images at low bitrates.

Video encoders such as MPEG \cite{mpeg} attempt to exploit extreme structural similarity (i.e., translational similarity) between adjacent video frames. However, apart from video data, exploitation of semantic similarities (i.e., similar high-level features such as objects, persons, etc.) remains a secondary priority in image compression.

\Section{Methods}
Our ``human image compression" setup circumvents modeling aspects of human visual perception by directly utilizing humans in a lossy image compression task. The setup involves two human participants, referred to as the describer and reconstructor, as presented in the Introduction.  

\begin{figure}[!htpb]
\centering
\includegraphics[width=6in]{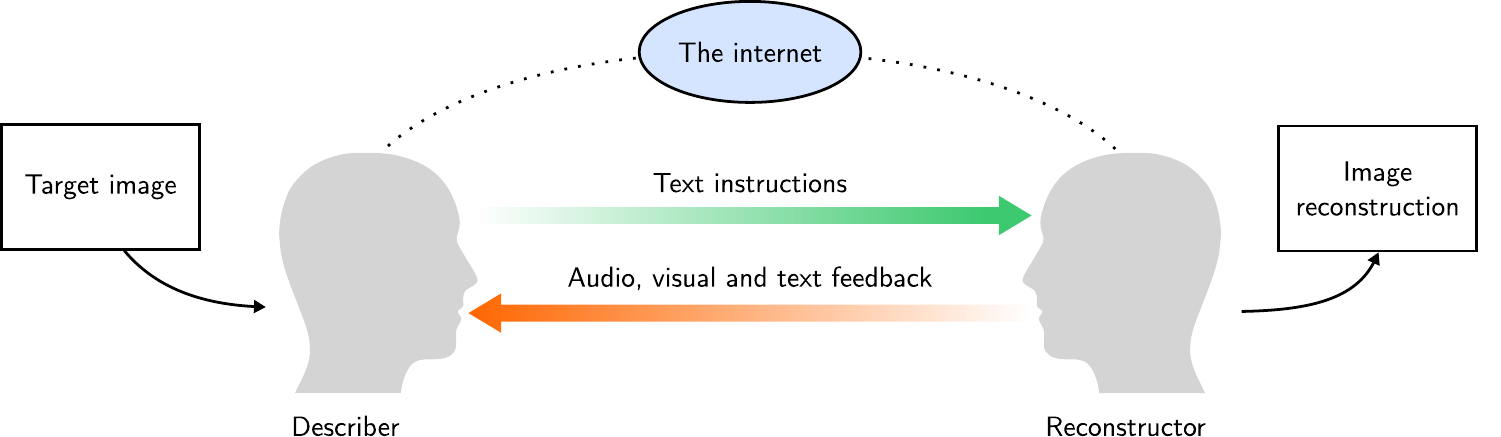}
\caption{\label{figure:setup} The human image compression setup. The describer fixes a target image, and attempts to describe it using text instructions, including URL links. The reconstructor attempts to reconstruct the image based on the text instructions. The describer is able to view the reconstruction, hear the reconstructor's voice and receive text feedback from the reconstructor. Both have access to the internet.}
\end{figure}
% \begin{figure}[htbp]
%     \centering
%     \includegraphics[width=\textwidth]{DCCTemplateLaTeX/Figures/block_diagram.pdf}
%     \caption{Block diagram showing the human compression process. The describer attempts to describe the image with URL links and text instructions. The describer can see the image as it is being reconstructed and can also hear the reconstructor's voice.}
%     \label{figure:reconstruction}
% \end{figure}

\newpage

\begin{figure}[!htpb]
\centering
\includegraphics[width=6in]{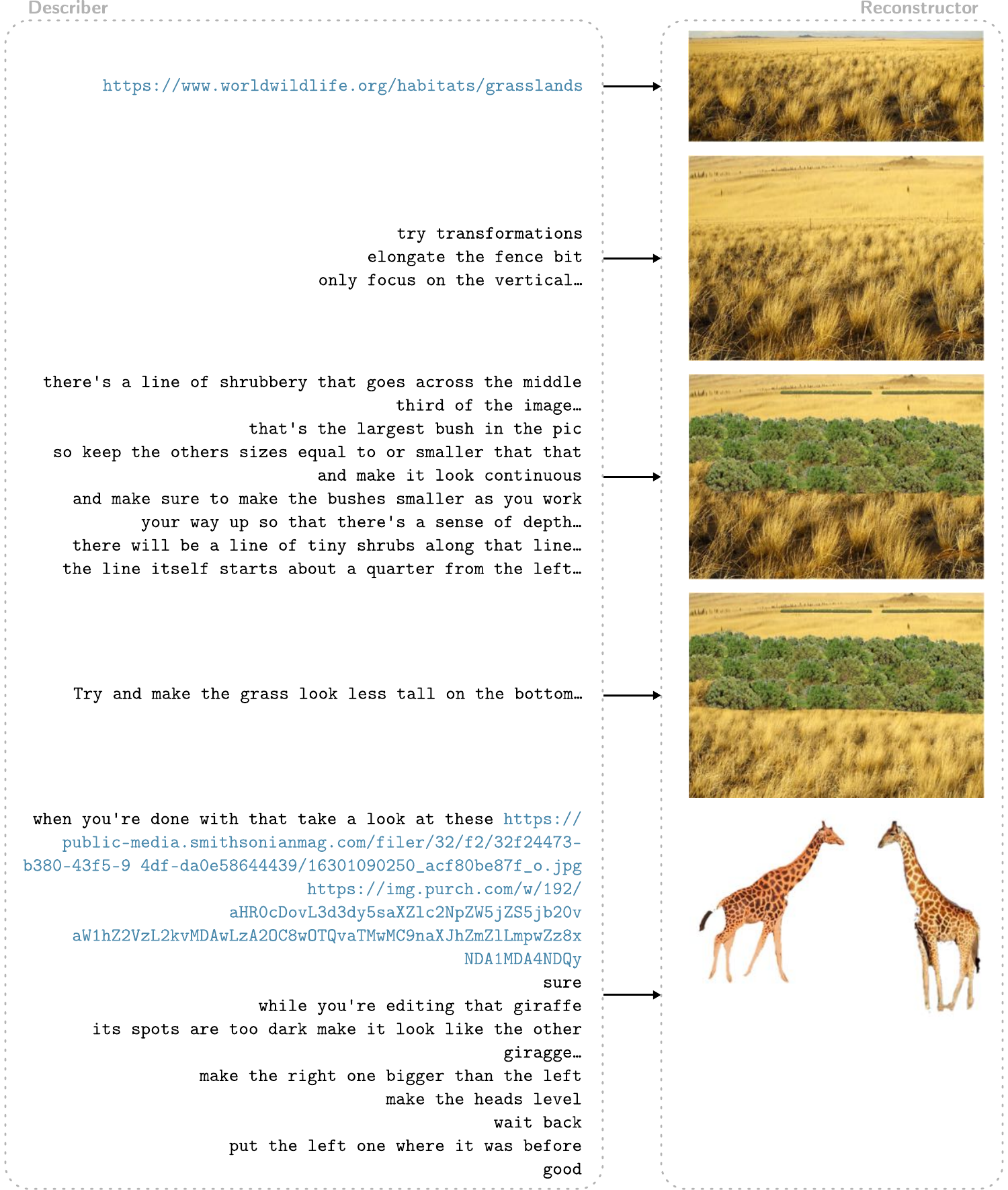}
\caption{\label{figure:reconstruction} Excerpts from a human compression session. A simplified excerpt of the reconstruction process for the giraffe image. The text on the left shows text communications sent from the describer to the reconstructor. Various stages of reconstruction are shown on the right, describing the background grass and bush, and the giraffes. Examples of internet links (blue) to publicly available images are also shown.}
\end{figure}

\newpage

For every input target image, the roles of the describer and reconstructor are as follows:
\begin{itemize}
    \item {\textbf{Describer}}: Analyzes/recognizes the input image and informs the reconstructor of the necessary steps to best recreate the target image. The describer communicates with the reconstructor only via real-time text chat and may view the reconstructed image in progress. They may also receive verbal communications from the reconstructor.
    \item {\textbf{Reconstructor}}: Interprets text instructions from the describer in order to produce a reconstruction of the original image. The reconstructor is not permitted to view the original image until the reconstruction is complete, but may communicate with the describer.
\end{itemize}
This human image compression paradigm combines two key aspects. First, it exploits human participants' preexisting competence in visual scene understanding and natural language use in order to elicit what information is prioritized in images. Second, it leverages public domain image data, thus avoiding the need to allocate additional disk space for visual information that is well-approximated by publicly available data. By permitting the exchange of natural language and pointers to publicly available images, our approach aims to characterize the limits of human-guided image compression under realistic expectations about the semantic knowledge and visual data that are shared between a sender and receiver \cite{clark1996using}. In principle, this empirical ``human-centric" approach may lead to the discovery of improved loss functions that respect human visual priorities to a greater degree than current compression techniques do.

In addition to descriptions and links to images, the describer may also send instructions for manipulating the linked images in order to create a satisfactory reconstruction. Altogether, our human compression scheme involves two streams of one-way communication: one text-based from the describer to the reconstructor, and one in any format from the reconstructor back to the describer. However, only the text from the describer to the reconstructor is considered to be the ``compressed" representation of the input image; any communication from the reconstructor to the describer is not counted towards the size of the final compressed representation of the image. To justify our accounting, we compare our human compression scheme to a machine implementation of compression.

In our experiments, the compression process involves communication between a describer and reconstructor which produces a text transcript, as well as a reconstructed image. The interactions between describer and reconstructor may be thought of as some sequence of instructions and actions. The describer's role is to issue instructions, and the reconstructor's job is to perform actions (e.g. reference image cropping, scaling, translation, etc.). Each instruction issued by the describer is based on their access to the target image as well as the previous action performed by the reconstructor, while the reconstructor performs actions in response to each instruction issued by the describer. The instruction-action process is repeated until the target image is reconstructed to the describer's satisfaction.

In lossy compression algorithm implementations, the compression process involves elements that function similarly to the human describer and human reconstructor. The elements that perform the description and reconstruction functions also interact much like their human counterparts do: machine ``instructions" are issued based on the target image and previous actions, and actions (e.g., prediction of the next block of pixels) are produced in response to the instructions received. The instruction-action process is repeated until the entire image is compressed, and generates a compressed representation, which is analogous to the transcript produced by the human compression setup.

However, unlike the human compression setup, in machine-implemented compression a reconstructed image is only produced when the decompression process is executed on the transcript. Notably, the decompression process may be thought of as \emph{identical} to the compression process, but with the describer replaced by the transcript. The stipulation of identicality necessitates that the actions performed by the reconstructor during decompression are identical to those performed during compression. In other words, the reconstructor must perform the same action in response to the same instruction, whether that instruction is from the describer or from a transcript. As a result, only the machine instructions flowing from the machine describer to the machine reconstructor are recorded, and the actions may be discarded from the transcript (cf. e.g., \cite{rateless_lossy}).

Of course, due to the large amount of variation in human cognition and behavior, it is unlikely for a human reconstructor to perform actions during the decompression process exactly as they did during the compression process. However, the fact remains that were the human reconstructor's actions able to be reproduced identically upon demand in response to a received instruction, then the text transcript containing only the describer's text instructions would suffice for creating an image reconstruction identical to that produced during the compression process. For this reason, in our human compression scheme we also consider only the describer's text to be counted towards the compressed representation of the target image. Furthermore, since a reconstructed image is produced in addition to a transcript of instructions, our setup may be thought of as simultaneous execution of both compression and decompression processes.

\SubSection{Implementation details}

The describer was provided an input image for compression, and a Skype call was initiated between the describer and reconstructor with the following restrictions. First, the describer could only communicate to the reconstructor through the inbuilt Skype text chat. The describer turned off their outgoing audio/video to avoid inadvertently leaking information to the reconstructor. Now, the reconstructor could communicate verbally with the describer through audio/video/text chat. Finally, the reconstructor could share partial, in-progress reconstructions with the describer in real time using Skype's screen share feature.

With these restrictions in place, the describer would begin to send a series of instructions for the reconstructor to attempt image reconstruction. Generally, the describer could send URL links to reference images that already exist on the internet, as well as specific text instructions for altering the image. A variety of image editing tasks could be sent, including: spatial translation of image elements, affine or perspective transformations, erasure or addition of certain objects in the image, enlargement of a portion of the image, compositing multiple images, etc. Figure \ref{figure:reconstruction} shows parts of the reconstruction process for the giraffe image.

When reconstruction had been completed to the level of the describer's satisfaction, the experiment was stopped. The Skype text transcript containing all instructions from the describer to the reconstructor was saved. Finally, the transcript was processed by removing timestamps and compressing it using the bzip2 \cite{bzip2} compressor. The bzip2-encoded Skype transcript represented the final compressed representation of the input image. The quality of image reconstruction can then be compared to that of a standard lossy image compressor.

\Section{Experiments}

\SubSection{Data Collection}
We first created a dataset of original images that are not publicly available on the web. The creation of original images prevents trivial encoding via an exact copy of a non-original picture. Original images were captured with a digital camera or smart-phone camera at high resolution. A wide variety of images (e.g., faces, landscapes, sketches, etc.) unknown to the describers and reconstructors were captured for the experiments. From these, we selected 13 diverse high-resolution images for our comparison experiments. The images and additional details are available in the appendix and at \url{https://compression.stanford.edu/human-compression}.

\SubSection{Experimental Setup}

We describe the experimental procedure for evaluating the quality of reconstructions by human compressors and WebP:

\begin{enumerate}
    \item \textbf{Human compression:} The input image is compressed and reconstructed by the human compression system using the procedure described in the Methods. The size (in bytes) of the compressed text instructions is recorded.
    \item \textbf{WebP compression:} The WebP compressor is used to lossily compress the input image to a size similar to that of the compressed human text instructions.
    \item \textbf{Quality evaluation:} The quality of WebP and human compressed images were compared using human scorers on the MTurk platform.
\end{enumerate}

\begin{figure}[!htpb]
\centering
\includegraphics[width=6in]{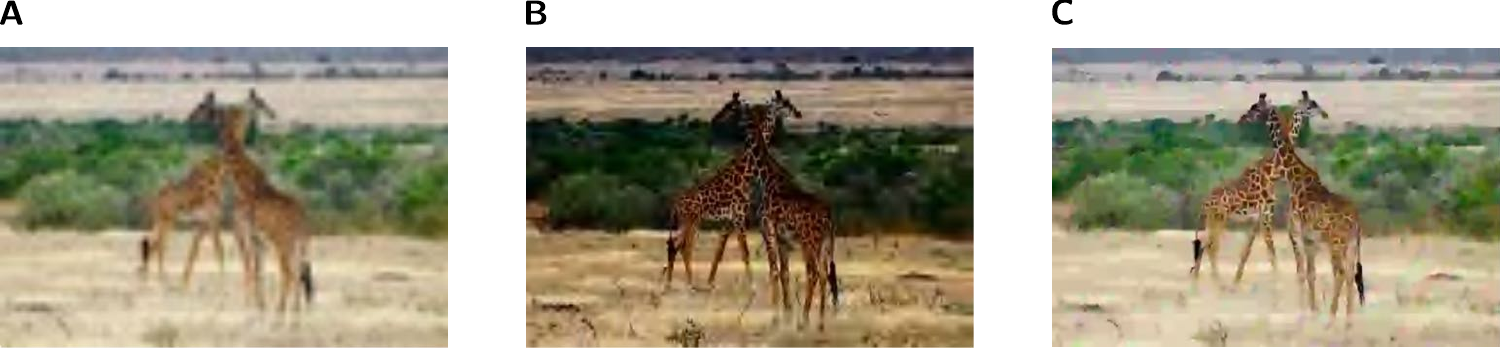}
\caption{\label{figure:jpeg_webp_human} A comparison of JPEG, JPEG2000 and WebP compression. The giraffe image compressed using (A) JPEG, (B) JPEG2000 and (C) WebP. All reconstructions are generated from a file size that is similar to that of the human-compressed giraffe file.}
\end{figure}

WebP \cite{WebP} is a relatively recent image compressor released by Google. We chose WebP as the reference compressor for comparing image reconstruction quality since WebP outperforms JPEG and JPEG2000 at the high compression levels achieved by the human compression scheme. This is illustrated in Figure \ref{figure:jpeg_webp_human}.

However, even when compressing images using WebP at the lowest allowed quality level (quality parameter set to 0), the compressed files were much larger than those of the human compressors. As a result, we first reduced the resolution of the images before compressing with WebP with quality parameter 0 in order to attain the target size, always erring on the side of the WebP file being larger than the compressed human text instructions.

\SubSection{Quality Evaluation using MTurk}
\begin{figure}[!htpb]
\centering
\includegraphics[width=6in]{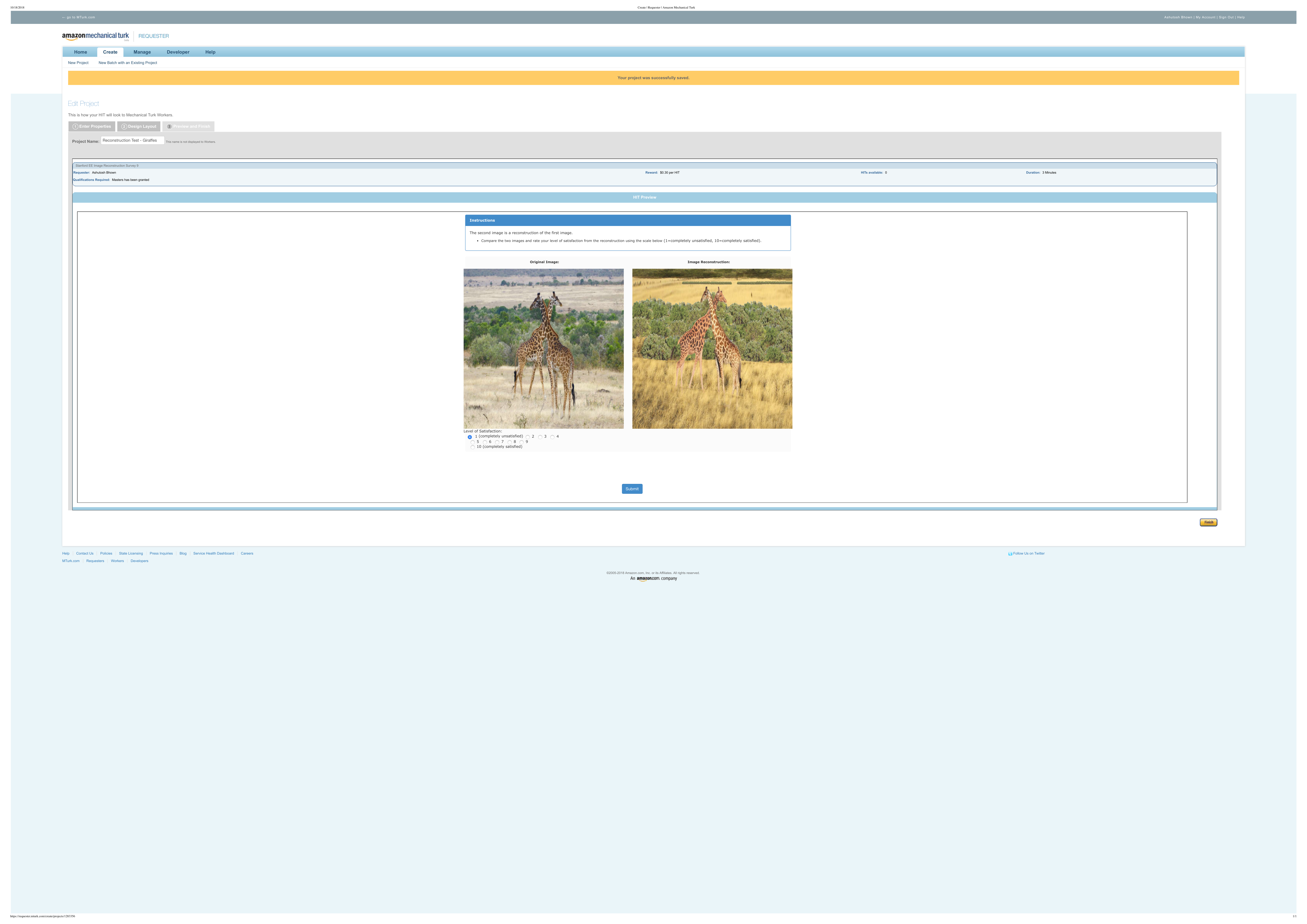}
\caption{\label{figure:mturk_compressed} Screen capture of the evaluation HIT that MTurk workers see in their web browser: a descriptive survey prompt at the top, the original image on the left, one image reconstruction on the right, and a rating scale below.}
\end{figure}

We compared the quality of compressed images using human scorers (workers) on Amazon Mechanical Turk (MTurk) \cite{mturk}, an online platform for conducting behavioral studies. For each image, we displayed the original image and a reconstructed image in a human intelligence task (HIT) which asked workers to rate the reconstruction on a scale of 1 to 10, according to their ``level of satisfaction'' with the reconstruction. For each HIT and for both types of reconstruction (human compression and WebP), we collected 100 survey responses and obtained summary statistics. Figure \ref{figure:mturk_compressed} shows a screenshot of the MTurk survey as seen by the workers. 

% \pagebreak

\Section{Results}

\begin{table}[!htbp]
\begin{center}
{
\renewcommand{\baselinestretch}{1}\footnotesize
\begin{tabular}{|c|c|c|c|c|c|c|c|}
\hline
\multirow{2}{*}{Image} & Original & Compressed chat & WebP size &\multicolumn{2}{c|}{Mean score} & \multicolumn{2}{|c|}{$\sigma_\mathrm{mean}$}\\
\cline{5-8}
 & size (KB) & size (KB) & (KB) & Human & WebP & Human & WebP\\
\hline
arch 	    &1119 & 3.805	&3.840& 4.04	& \textbf{5.1}	& 0.23	& 0.21\\
balloon	    & 92 & 	1.951&	2.036& \textbf{6.22}	& 5.45	& 0.23	& 0.25\\
beachbridge	&3263	&4.604&	4.676& \textbf{4.34}	& 3.92	& 0.23&	0.22\\
eiffeltower	&2245&	4.363&	4.394& \textbf{5.98}	& 5.77	& 0.22&	0.22\\
face	    &1885	&2.649&	2.762& 2.95	& \textbf{5.47}	& 0.19	&0.20\\
fire	    &4270&	2.407	&2.454& \textbf{6.74}	& 5.09	& 0.23	& 0.23\\
giraffe	    &5256	&3.107	&3.144& \textbf{6.28}	& 4.48	& 0.24	&0.21\\
guitarman	&1648	&2.713&	2.730& \textbf{4.88}	& 4.07	& 0.26	& 0.20\\
intersection&3751	&3.157&	3.238& \textbf{6.8}	& 4.15	& 0.19	&0.22\\
rockwall	&4205	&6.613&	6.674& 4.41	& \textbf{4.85}	& 0.23 &	0.23\\
sunsetlake	&1505	&4.077&	4.088& \textbf{5.08}	& 4.82	& 0.23&	0.23\\
train	    &3445&	1.948&	2.024& \textbf{6.85}	& 3.62	& 0.23	& 0.21\\
wolfsketch	&1914&	0.869&	0.922& \textbf{8.25}	& 3.46	& 0.20	&0.19\\
\hline
\end{tabular}}
\caption{\label{tab:main_result}%
Original image size and compressed sizes along with mean MTurk scores for human and WebP reconstructions. Best results are boldfaced, and standard error of the mean ($\sigma_\mathrm{mean}$, sample size 100 for each compression method) for all scores are shown.}
\end{center}
\end{table}

\begin{figure}[!htpb]
\centering
\includegraphics[width=6in]{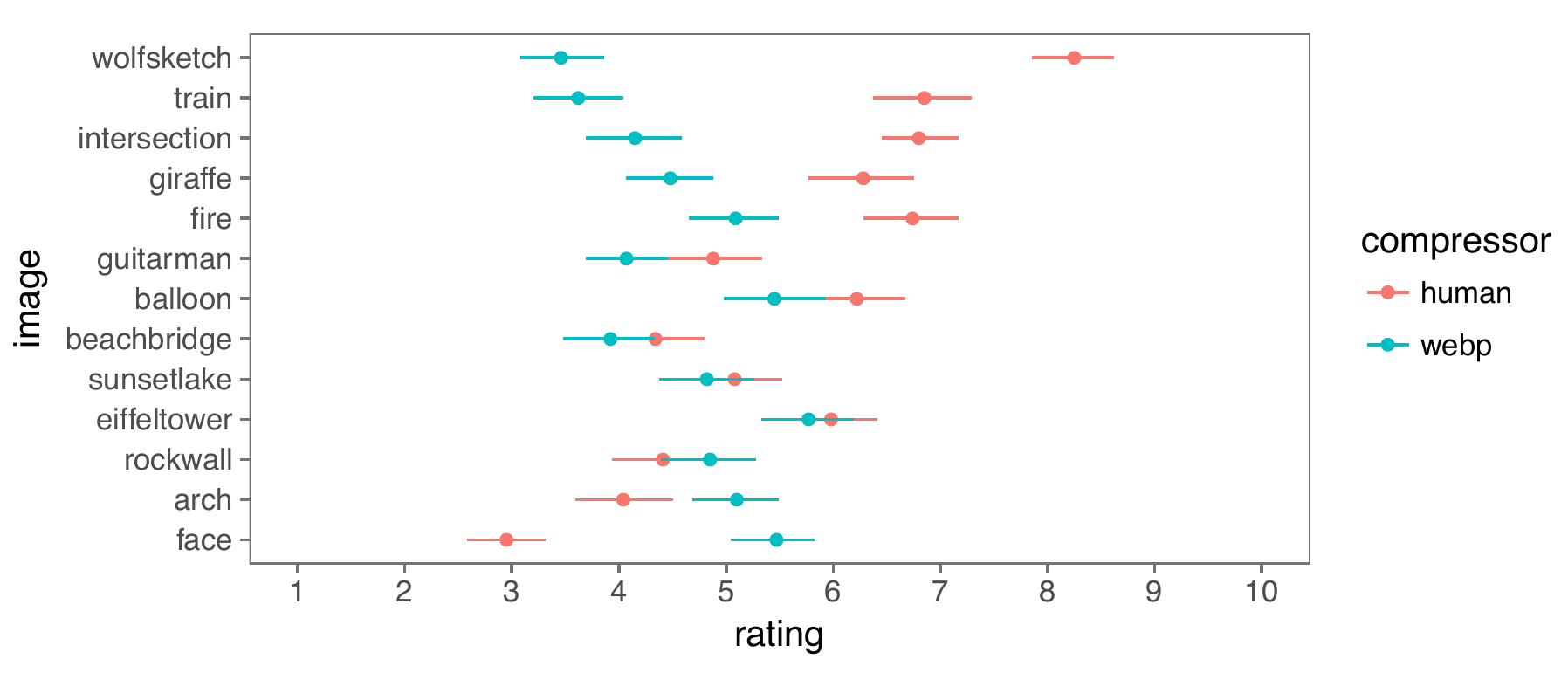}
\caption{\label{figure:confidence_intervals} Mean ratings and 95\% confidence intervals for human and WebP compression, ordered by the difference in mean ratings between the two compressor types.}
\end{figure}

Table \ref{tab:main_result} shows the mean ratings given by MTurk workers to human and WebP compressed reconstructions of the 13 high-resolution images. Figure \ref{figure:confidence_intervals} visualizes the distribution of these ratings with corresponding 95\% confidence intervals which were obtained via bootstrap resampling 1000 times \cite{efron1986bootstrap}.

We fit these ratings with a linear mixed-effects regression model predicting rating from compressor type (human vs. WebP), with random intercepts for different images and human scorers.
This analysis revealed a marginal advantage for human image reconstructions relative to WebP compressed images, which were rated 0.984 points higher on average (t = 1.82, p = 0.090).
This suggests that while the current study may be underpowered to detect a statistically reliable effect across images, larger studies containing more images may reveal more consistent differences in reconstruction quality accomplished by each compression method. 
\begin{figure}[!htpb]
\centering
\includegraphics[width=6in]{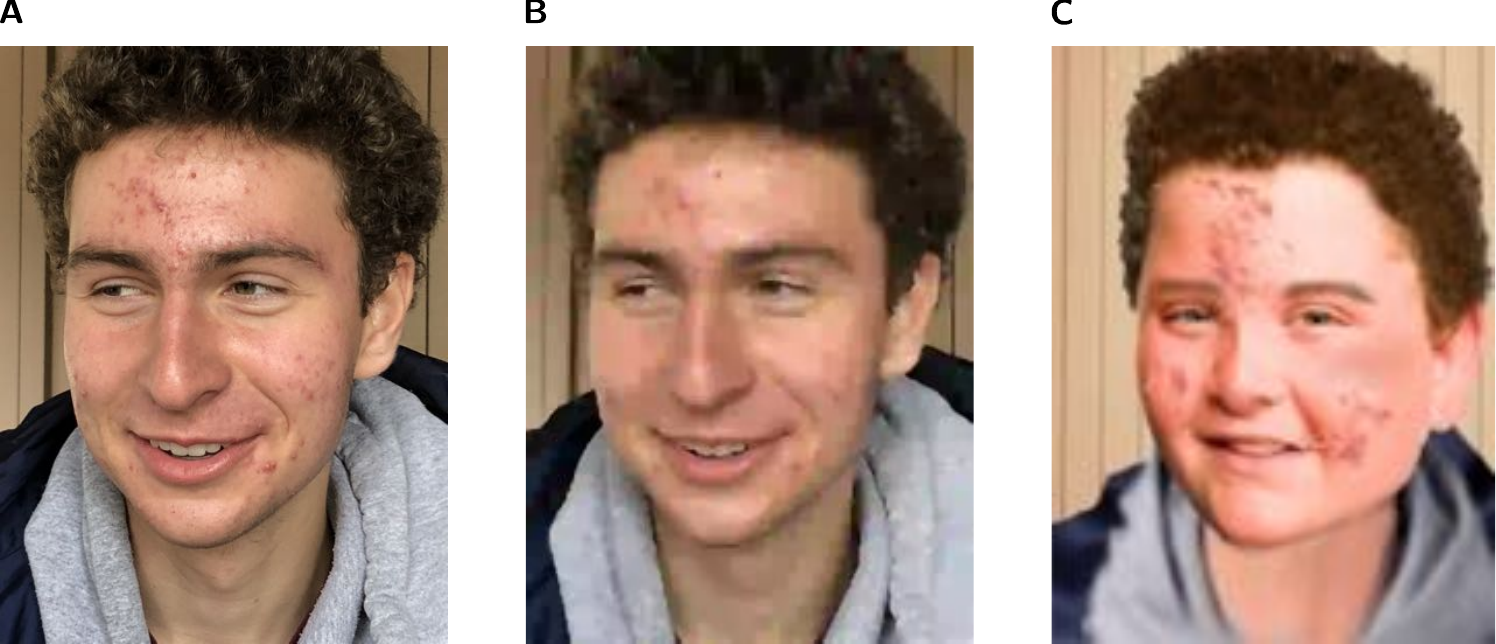}
\caption{\label{figure:face_results} Face image results. (A) Original face image with (B) WebP and (C) human reconstructions.}
\end{figure}

\begin{figure}[!htpb]
\centering
\includegraphics[width=6in]{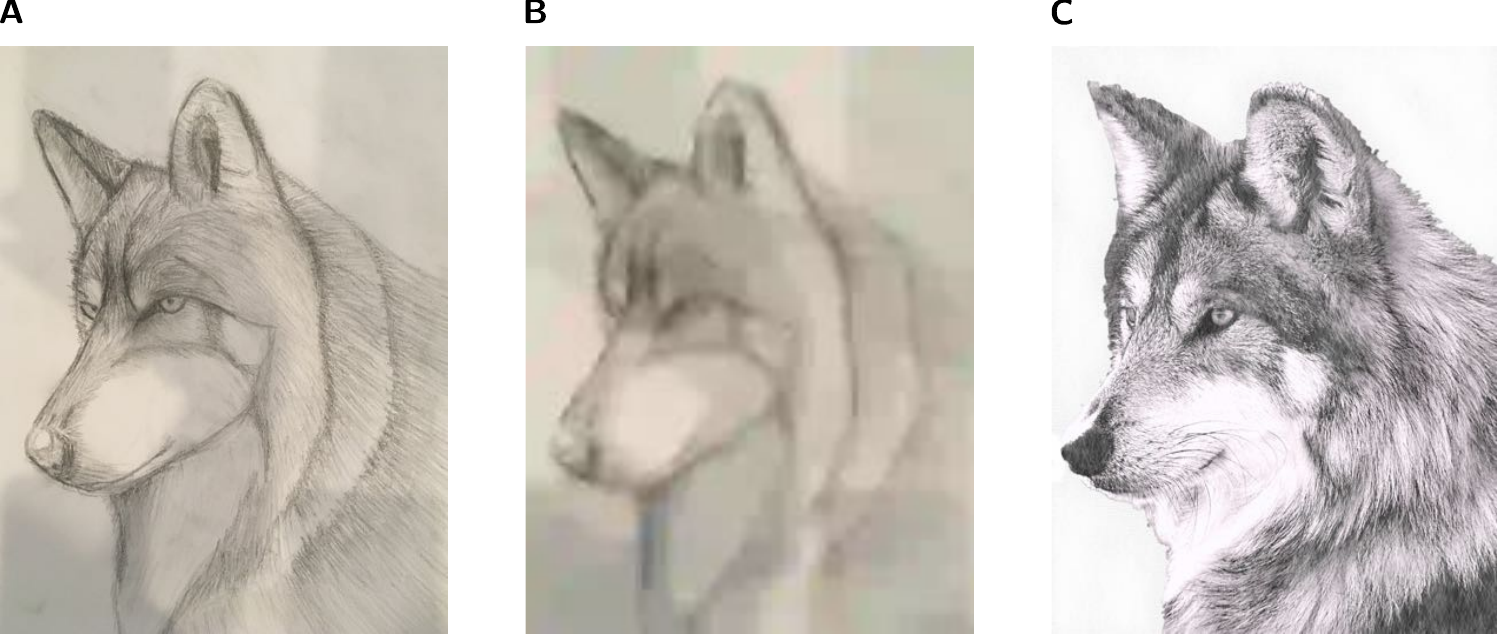}
\caption{\label{figure:wolfsketch_results} Wolfsketch image results. (A) Original wolfsketch image with (B) WebP and (C) human reconstructions.}
\end{figure}

Importantly, our study also revealed a large degree of variation in both the absolute ratings given to different images, as well as the magnitude of the difference between human and WebP reconstructions. For some images, the human reconstructions were judged to be clearly higher in quality relative to the WebP compressed images (see Figures \ref{figure:giraffe_intro}, \ref{figure:face_results}, \ref{figure:wolfsketch_results}), while still achieving high compression ratios ranging from around 100$\times$ to 1000$\times$. For the giraffe image (Figure \ref{figure:giraffe_intro}), we suspect that human reconstruction achieved a better rating than WebP because human scorers give more priority to image sharpness over accuracy. In contrast, for the face image (Figure \ref{figure:face_results}) human compression achieved a significantly lower quality score than WebP. We speculate that this is because facial identity is more important than individual semantic features of the image (such as the presence of facial blemishes). On the other hand, humans achieve a much better score than WebP for the wolfsketch image (Figure \ref{figure:wolfsketch_results}), perhaps because human scorers are not as sensitive to differences in wolf identity. We also observed that human compression achieved better compression ratios and MTurk scores when semantically similar images were publicly available. This was the case for images of famous monuments such as the eiffeltower image, and for the intersection image where Google Street View provided images of similar road intersections.

\Section{Discussion \& Conclusion}

We designed an experiment to better understand the potential for improving lossy image compression based on a human-centric loss. In the context of this two-player image reconstruction game, human participants played the roles of describer and reconstructor and generated compressed versions of 13 diverse images of landscapes, portraits, animals and urban settings. We evaluated the quality of human compression by comparing their reconstructions with those generated from WebP compression. For several of the images, the human reconstructions were preferred to the WebP reconstructions (e.g., wolf, train). For those images, the human compression process was better at identifying and preserving image properties that were relevant to human scorers. However, for a number of images the WebP reconstructions were preferred over the human reconstructions (e.g., face, arch). For those images, it appears that the publicly available image data and reconstruction interface may not have been sufficient to preserve the attributes that people considered to be most important. We plan to follow up these preliminary observations with a larger human reconstruction study containing more images. A wider breadth of test images should provide a more precise estimate of the relative quality of human reconstructions, as well as a better understanding of how the type of semantic information in an image affects how well a reconstruction can be achieved using simple operations on publicly available image data. %Overall, our findings suggest that there may be substantial room for improving image compression using human-centric loss functions combined with growing public image databases.

The human compression scheme is able to exploit semantically similar images quite effectively during compression. However, most popular compressors do not appear to take advantage of this rich public resource. Our experiment suggests that effective utilization of semantically and structurally similar images (or parts of images) can dramatically improve compression ratios. This is particularly relevant today, when images can be easily found using image search tools such as the one offered freely by Google. 

While the human compression framework is useful as an exploratory tool, it is clearly not practical due to its labor-intensive nature. We did not strive to optimize our protocols in any way, and we could have undoubtedly achieved substantially better compression and reconstruction scores had we done so. Notably, each of the image reconstructions took a few hours to complete. Furthermore, redundancies in English language resulted in sub-optimal compression, even though this is partly resolved by the use of bzip2. Our drawing skills, use of rudimentary software for image editing, inefficiencies due to occasional misunderstandings of describer instructions by the reconstructor, and difficulty in manually searching for similar images all contributed to transcript size. Improvements on any of these fronts would further result in improved image reconstruction quality.

We plan to use the insights obtained from this work to build an image compressor that is both optimized for human perception loss and able to utilize side information in the form of publicly available databases. We look to the work in \cite{perceptual}, which trains a neural network to predict human scores, as a strategy for training machine-based compressors for the human perception loss. We also expect to take advantage of reverse image search tools in order to better utilize side information. We believe these techniques will be key to significantly improved lossy image compression.

We also intend to further explore the theoretical limits of information transfer using both state-of-the art image compressors as well as our human-inspired image compression setup. Our work was inspired in part by Claude Shannon's 1951 paper \cite{shannon1951prediction}, where humans were used to establish an upper bound on the fundamental limit of English language compression. At the time, humans were better compressors than any practically implementable algorithm, and the paper motivated subsequent developments in text compression to match and eventually surpass the 2.3 bits/symbol shown to be achievable by human compressors.
Towards this end, in future experiments we plan to generate and score human reconstructions at several compression levels for each image, and to compare the resultant reconstruction versus quality curves with those achieved by WebP. 
This approach would provide a more comprehensive characterization of the fundamental tradeoff between compression rate and reconstruction quality \cite{shannon1948mathematical} for both state-of-the-art compressors and human compressors, calibrated to the same evaluation metric. 
The results from such a study may guide development of lossy image compression algorithms that will achieve and eventually surpass human performance.

\Section{Acknowledgement}
We thank Meltem Tolunay, Yihui Quek, Jay Mardia, Yanjun Han, Dmitri Pavlichin and Ariana Mann for fruitful discussions.  We also thank Debargha Mukherjee for his helpful comments on the manuscript. We thank Lucas Washburn for permitting us to take his photo and use it in our experiments.  We also thank the NSF Center for the Science of Information, NIH, the Stanford Compression Forum and Google for funding various parts of this project.

\Section{References}

%\bibliographystyle{IEEEbib}
%\bibliography{refs}

\clearpage

\Section{Appendix}
\SubSection{Additional details}
Table \ref{tab:image_details}% and \ref{tab:mturk_detailed} 
contains additional details about the images and the mechanical turk experiments.
\begin{table}[htbp]
\begin{center}
{
\renewcommand{\baselinestretch}{1}\footnotesize
\begin{tabular}{|c|c|c|c|c|c|}
\hline
\multirow{2}{*}{Image} & Original & WebP & Original & Compressed chat & WebP size \\
& resolution & resolution & size (KB) & size (KB) & (KB) \\ 
\hline
arch&	1762 $\times$ 2286	&506 $\times$ 656	&1119 & 3.805	&3.840\\
balloon&	1024 $\times$ 683	&630 $\times$ 420& 92 & 	1.951&	2.036\\
beachbridge&	4032 $\times$ 3024	&500 $\times$ 375 &3263	&4.604&	4.676\\
eiffeltower&	2448 $\times$ 3264	&492 $\times$ 656&2245&	4.363&	4.394\\
face&	3024 $\times$ 4032	&435 $\times$ 580&1885	&2.649&	2.762\\
fire&	3036 $\times$ 4048	&375 $\times$ 500 &4270&	2.407	&2.454\\
giraffe&	5472 $\times$ 3648	&528 $\times$ 352 &5256	&3.107	&3.144\\
guitarman&	1136 $\times$ 640	&550 $\times$ 310 &1648	&2.713&	2.730\\
intersection&	3024 $\times$ 4032&	450 $\times$ 600 &3751	&3.157&	3.238\\
rockwall&	3036 $\times$ 4048	&531 $\times$ 708&4205	&6.613&	6.674\\
sunsetlake&	3264 $\times$ 2448	&1148 $\times$ 861&1505	&4.077&	4.088\\
train&	4032 $\times$ 3024	&340 $\times$ 255&3445&	1.948&	2.024\\
wolfsketch&	2698 $\times$ 3539	&290 $\times$ 380&1914&	0.869&	0.922\\
\hline
\end{tabular}}
\caption{%
\small Resolution and original/compressed size for the images. Chat transcripts were compressed with bzip2. WebP resolution was reduced till the file size just exceeded the compressed chat transcript size, keeping quality parameter 0 and aspect ratio fixed.}
\label{tab:image_details}
\end{center}
\end{table}

% \begin{table}[htbp]
% \begin{center}
% {
% \renewcommand{\baselinestretch}{1}\footnotesize
% \begin{tabular}{|c|c|c|c|c|c|c|}
% \hline
% \multirow{2}{*}{Image} &\multicolumn{2}{c|}{Mean score} & \multicolumn{2}{|c|}{Median score} & \multicolumn{2}{|c|}{Standard deviation} \\
% \cline{2-7}
%  & Human & WebP& Human & WebP& Human & WebP\\
% \hline
% arch 	    & 4.04	& \textbf{5.1}	& 3	& \textbf{5}&2.27	&2.11\\
% balloon	    & \textbf{6.22}	& 5.45	& \textbf{7}	& 6&2.3	&2.54\\
% beachbridge	& \textbf{4.34}	& 3.92	& 4	& 4 &2.27&	2.17\\
% eiffeltower	& \textbf{5.98}	& 5.77	& 6	& 6&2.2&	2.15\\
% face	    & 2.95	& \textbf{5.47}	& 3	& \textbf{6}&1.87	&2.01\\
% fire	    & \textbf{6.74}	& 5.09	& \textbf{7}	& 5&2.31	&2.25\\
% giraffe	    & \textbf{6.28}	& 4.48	& \textbf{7}	& 4&2.37	&2.08\\
% guitarman	& \textbf{4.88}	& 4.07	& \textbf{5}	& 4&	2.55	&2.03\\
% intersection& \textbf{6.8}	& 4.15	& \textbf{7}	& 4&	1.9	&2.17\\
% rockwall	& 4.41	& \textbf{4.85}	& 4	& \textbf{5}	&2.33&	2.27\\
% sunsetlake	& \textbf{5.08}	& 4.82	& 5	& 5&2.33&	2.34\\
% train	    & \textbf{6.85}	& 3.62	& \textbf{7}	& 3&2.3	&2.1\\
% wolfsketch	& \textbf{8.25}	& 3.46	& \textbf{9}	& 3&2.03	&1.94\\
% \hline
% \end{tabular}}
% \caption{%
% \small
% Mean, median and standard deviation of MTurk scores for human and WebP reconstructions. Best result for each image is boldfaced.}
% \label{tab:mturk_detailed}
% \end{center}
% \end{table}

\SubSection{Images}
This section contains all 13 original images along with their WebP and human reconstructions.
\begin{figure}[!htpb]
\centering
\includegraphics[width=6in]{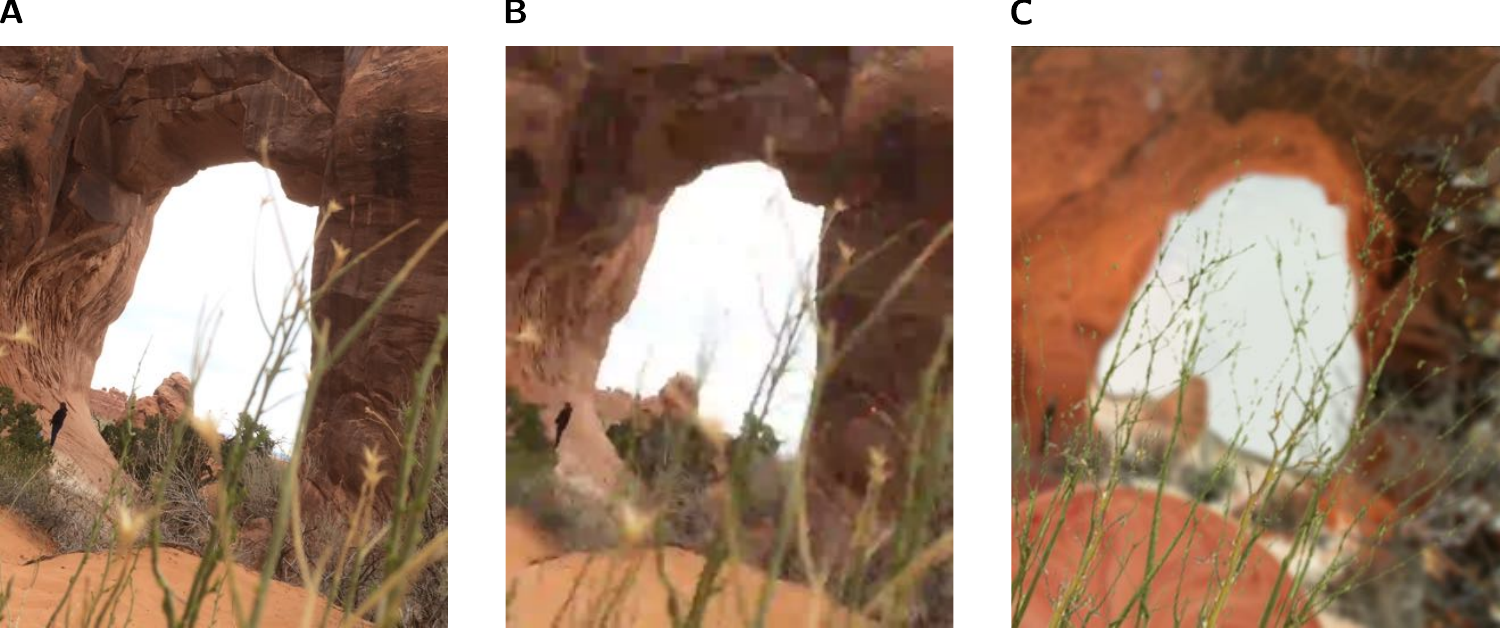}
\caption[Arch image results]{\label{figure:arch_results} (A) Original arch image with (B) WebP and (C) human reconstructions.}
\end{figure}

\begin{figure}[!htpb]
\centering
\includegraphics[width=6in]{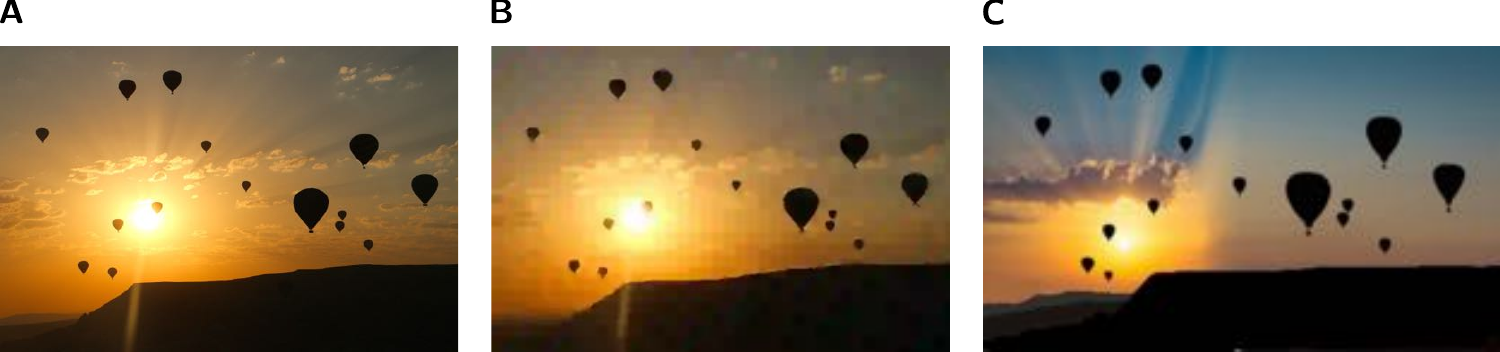}
\caption[Balloon image results]{\label{figure:balloon_results} (A) Original balloon image with (B) WebP and (C) human reconstructions.}
\end{figure}

\begin{figure}[!htpb]
\centering
\includegraphics[width=6in]{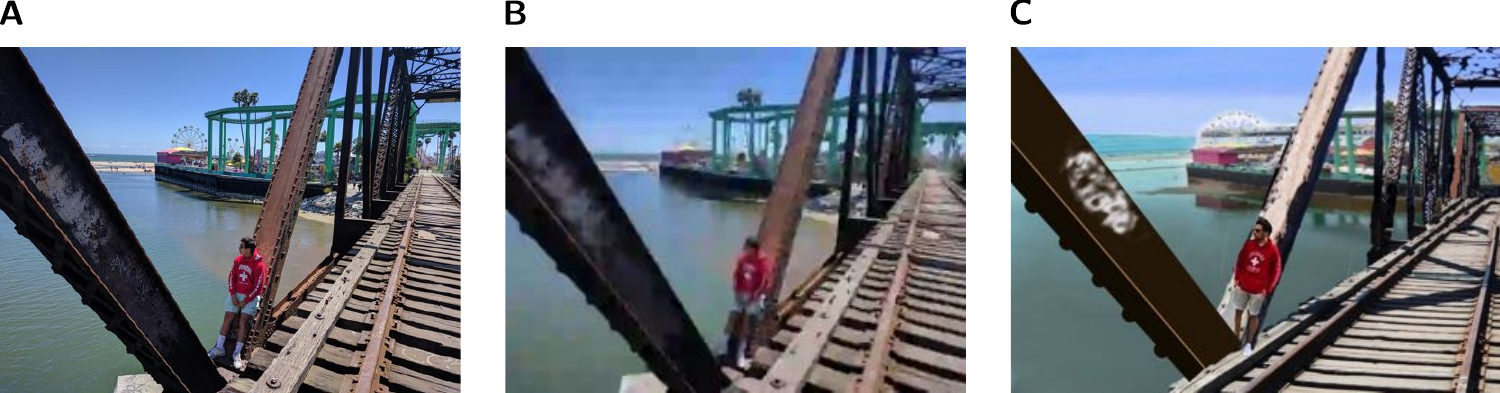}
\caption[Beachbridge image results]{\label{figure:beachbridge_results} (A) Original beachbridge image with (B) WebP and (C) human reconstructions.}
\end{figure}

\begin{figure}[!htpb]
\centering
\includegraphics[width=6in]{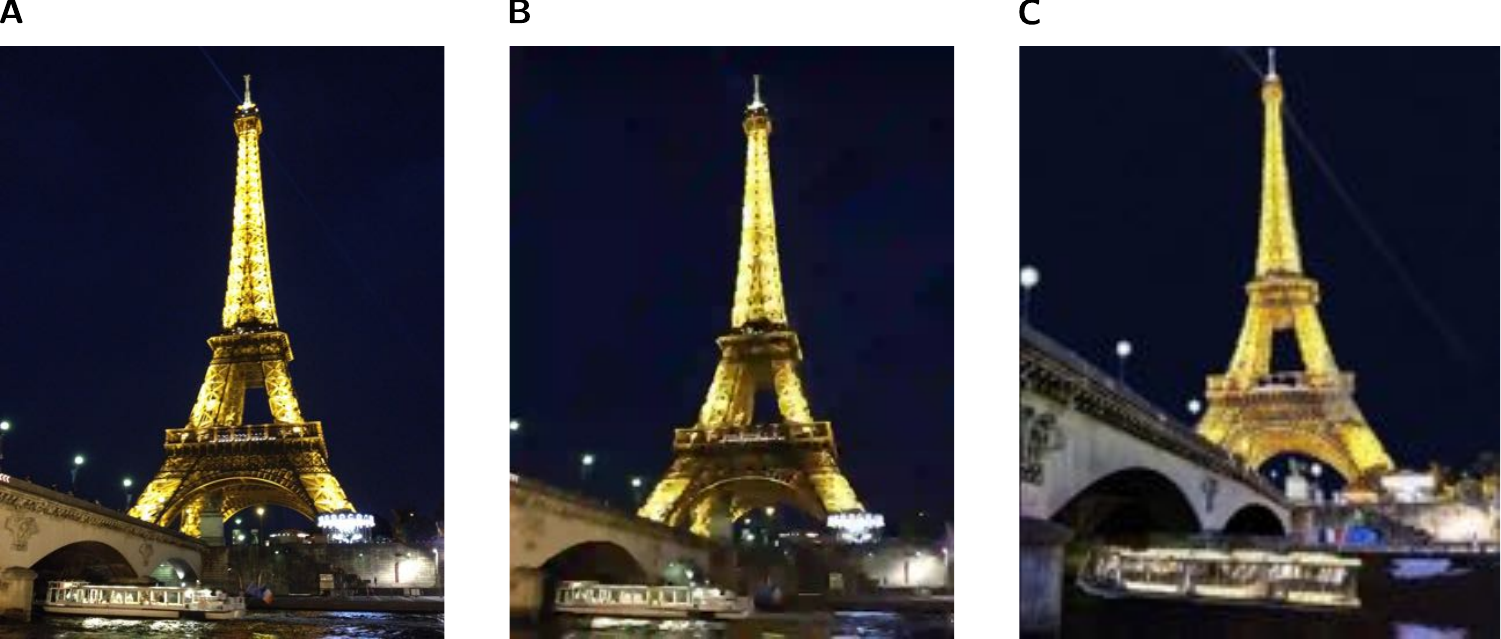}
\caption[Eiffeltower image results]{\label{figure:eiffeltower_results} (A) Original eiffeltower image with (B) WebP and (C) human reconstructions.}
\end{figure}

\begin{figure}[!htpb]
\centering
\includegraphics[width=6in]{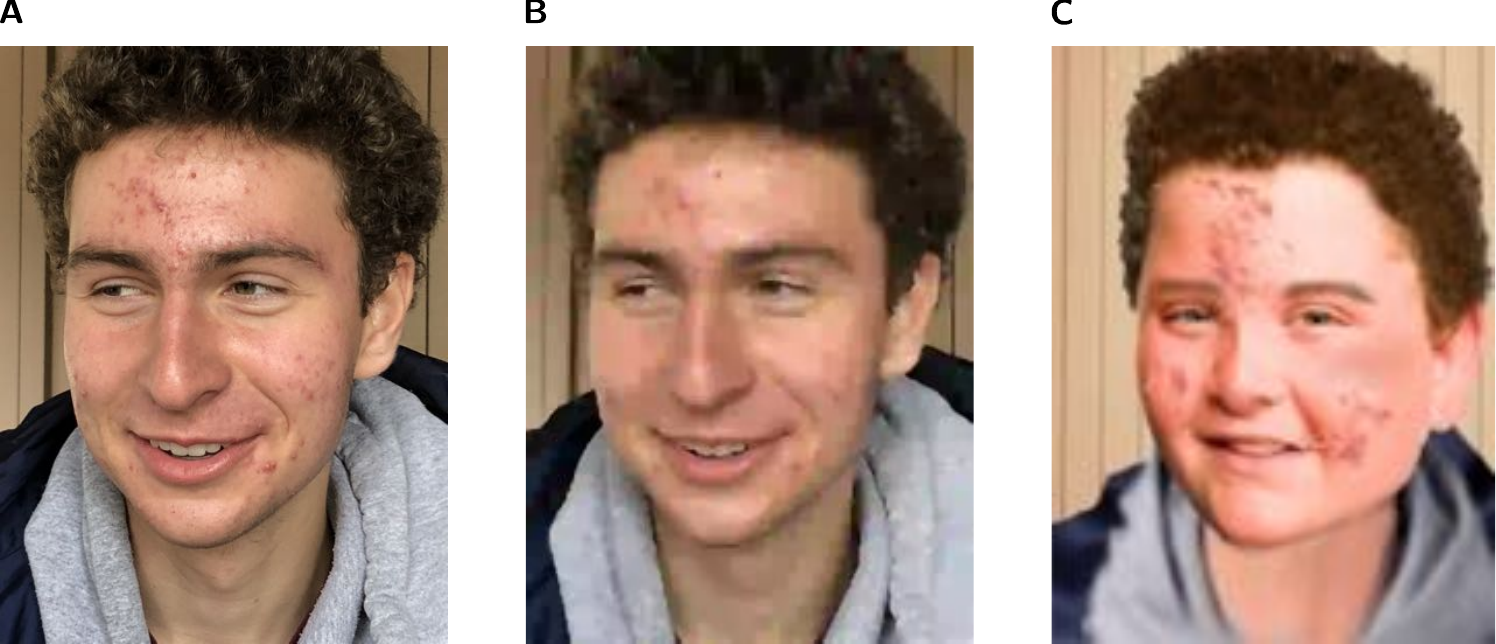}
\caption[Face image results]{\label{figure:face_results_appendix} (A) Original face image with (B) WebP and (C) human reconstructions.}
\end{figure}

\begin{figure}[!htpb]
\centering
\includegraphics[width=6in]{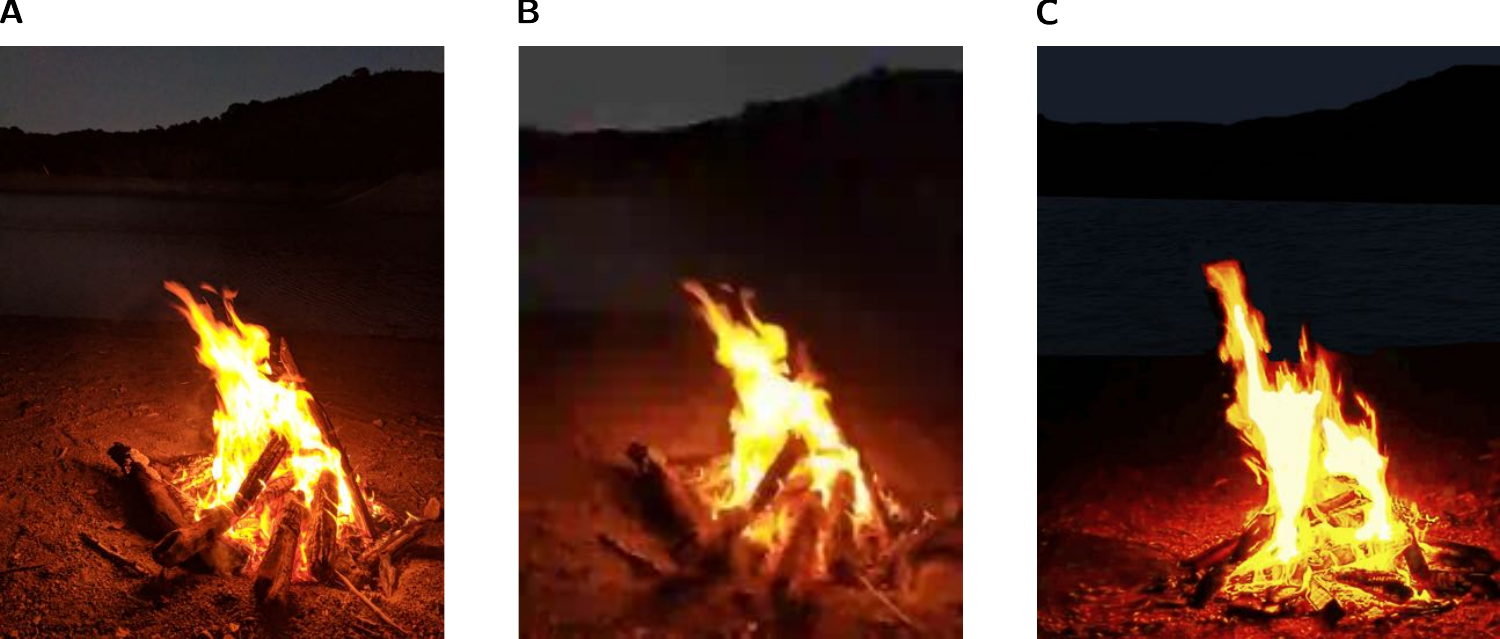}
\caption[Fire image results]{\label{figure:fire_results} (A) Original fire image with (B) WebP and (C) human reconstructions.}
\end{figure}

\begin{figure}[!htpb]
\centering
\includegraphics[width=6in]{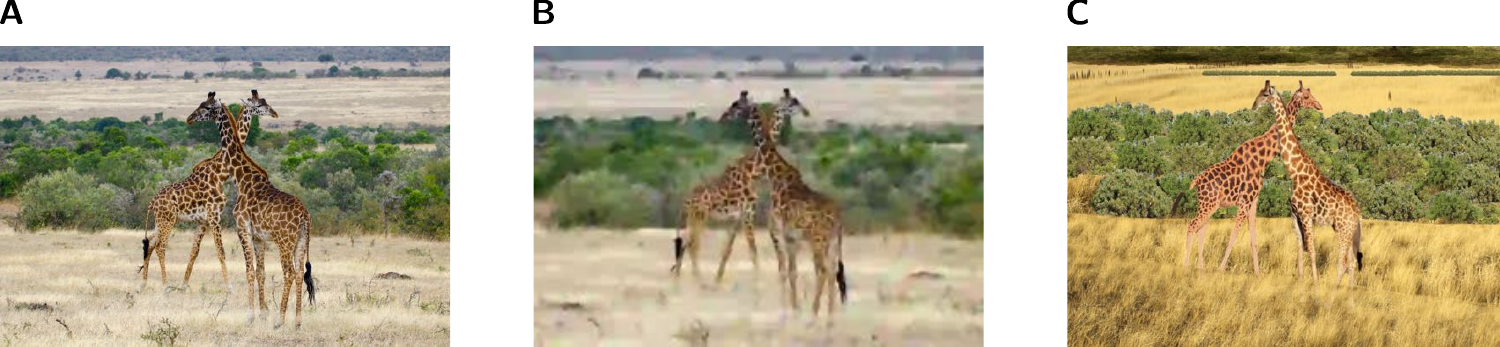}
\caption[Giraffe image results]{\label{figure:giraffe_results_appendix} (A) Original giraffe image with (B) WebP and (C) human reconstructions.}
\end{figure}

\begin{figure}[!htpb]
\centering
\includegraphics[width=6in]{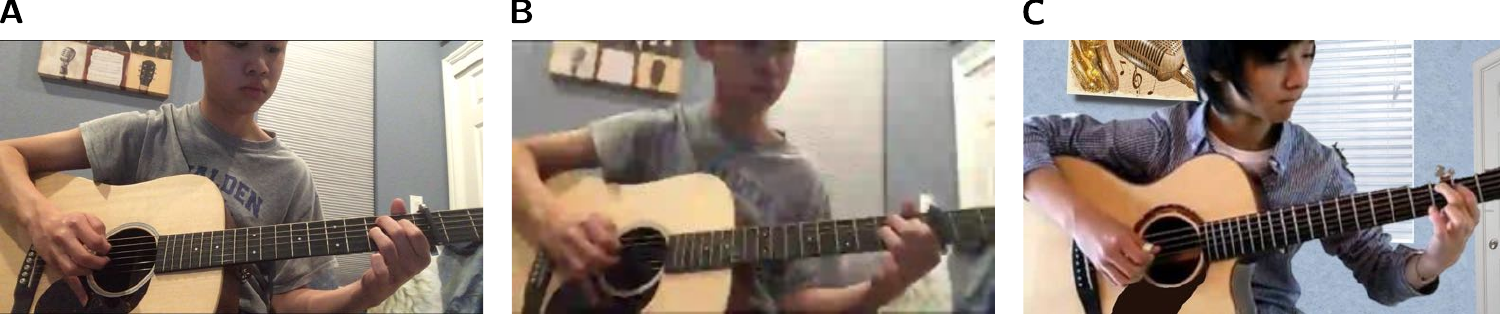}
\caption[Guitarman image results]{\label{figure:guitarman_results} (A) Original guitarman image with (B) WebP and (C) human reconstructions.}
\end{figure}

\begin{figure}[!htpb]
\centering
\includegraphics[width=6in]{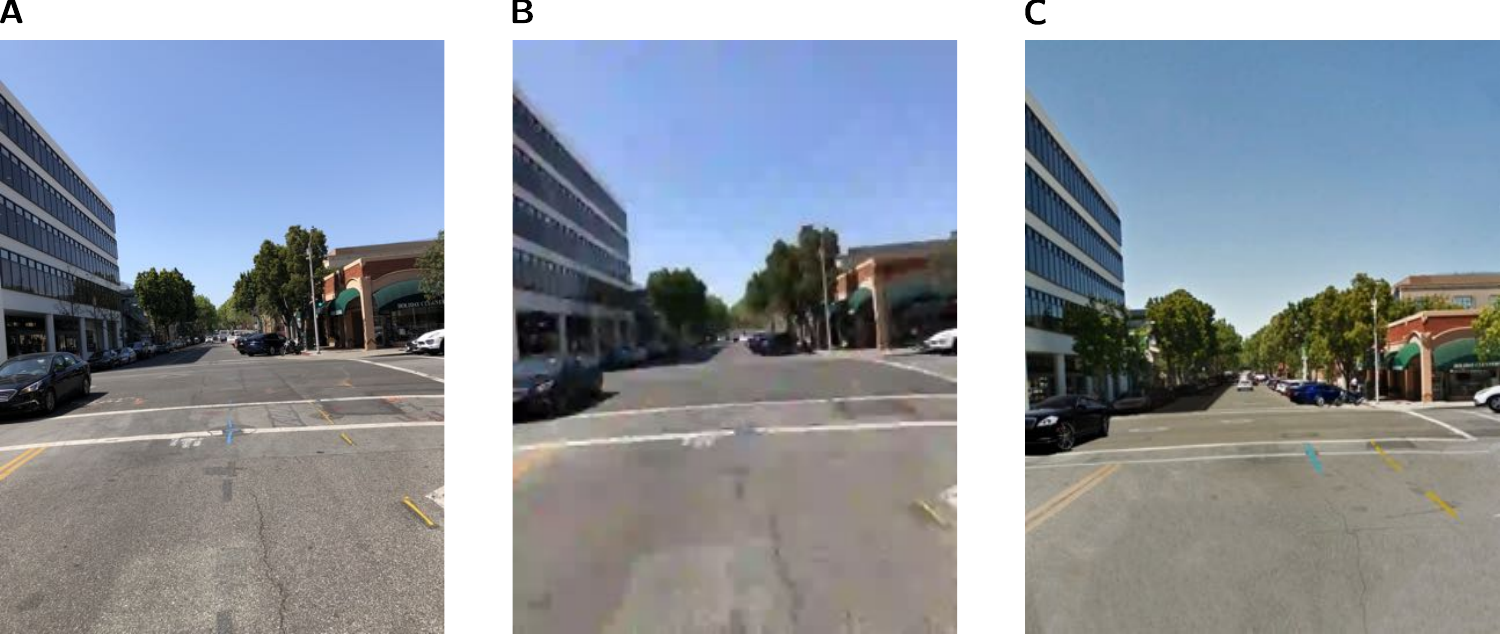}
\caption[Intersection image results]{\label{figure:intersection_results} (A) Original intersection image with (B) WebP and (C) human reconstructions.}
\end{figure}

\begin{figure}[!htpb]
\centering
\includegraphics[width=6in]{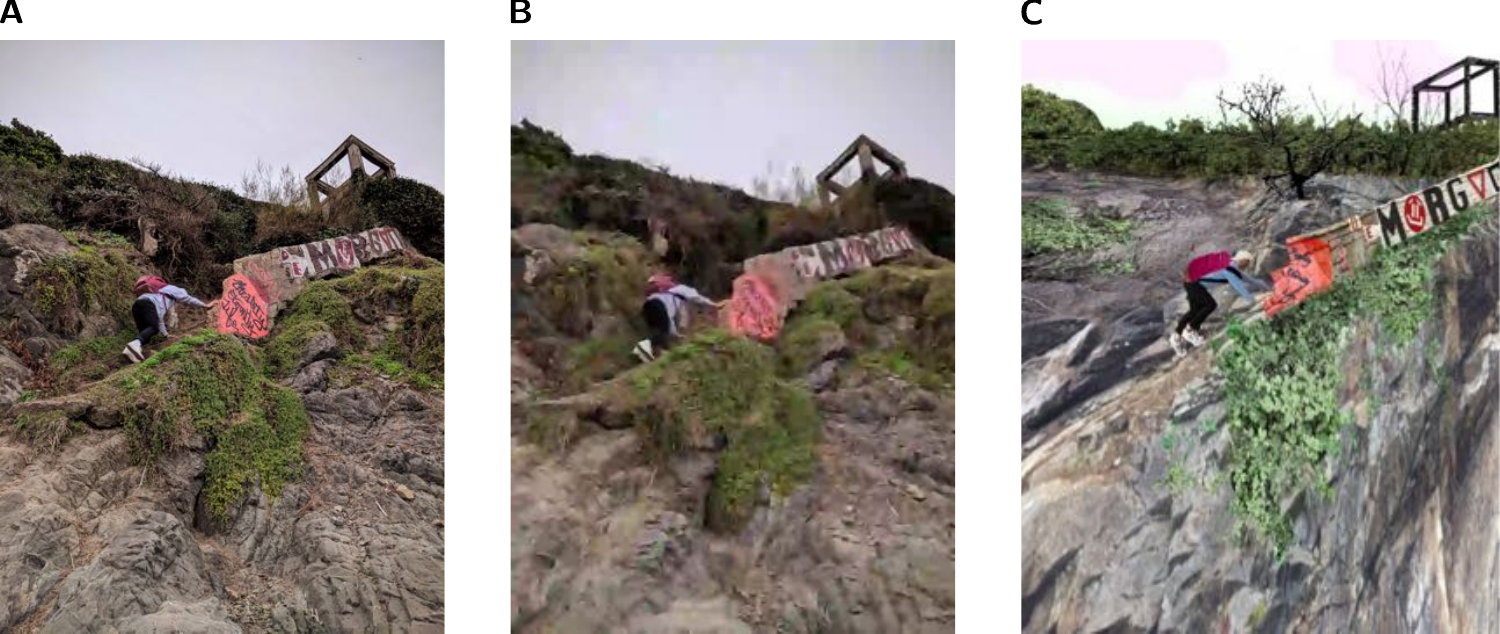}
\caption[Rockwall image results]{\label{figure:rockwall_results} (A) Original rockwall image with (B) WebP and (C) human reconstructions.}
\end{figure}

\begin{figure}[!htpb]
\centering
\includegraphics[width=6in]{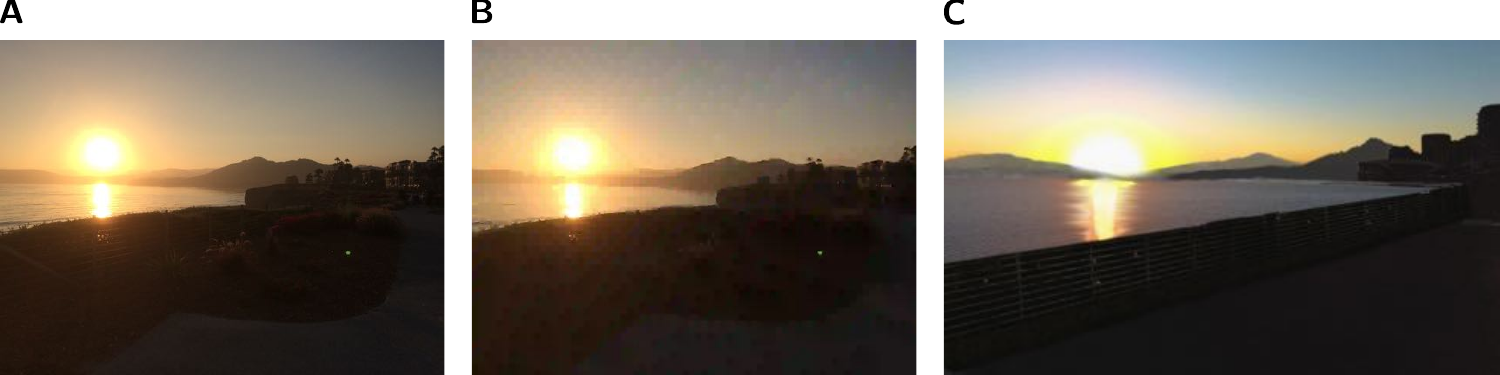}
\caption[Sunsetlake image results]{\label{figure:sunsetlake_results} (A) Original sunsetlake image with (B) WebP and (C) human reconstructions.}
\end{figure}

\begin{figure}[!htpb]
\centering
\includegraphics[width=6in]{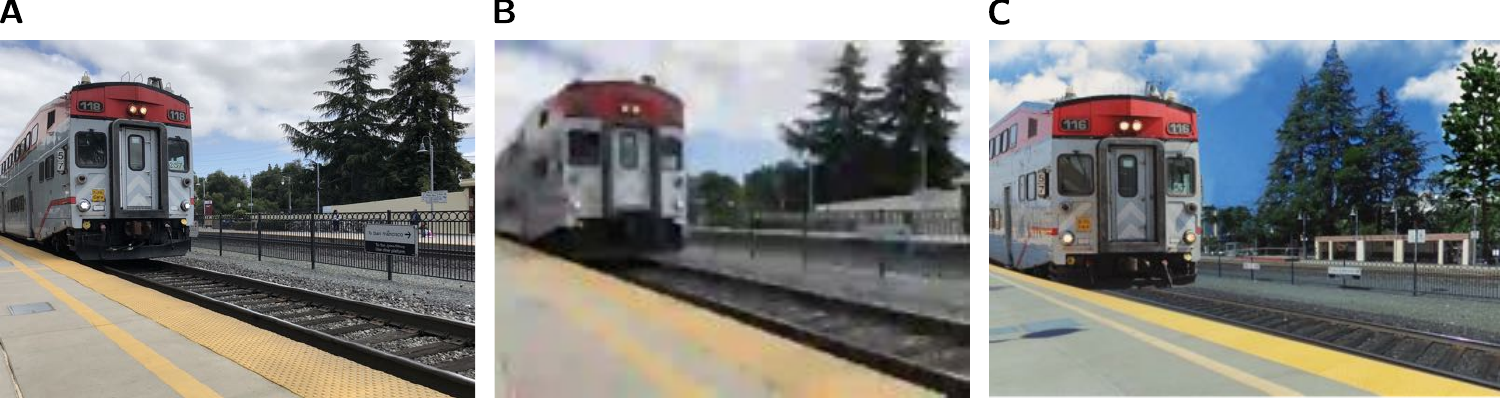}
\caption[Train image results]{\label{figure:train_results} (A) Original train image with (B) WebP and (C) human reconstructions.}
\end{figure}

\begin{figure}[!htpb]
\centering
\includegraphics[width=6in]{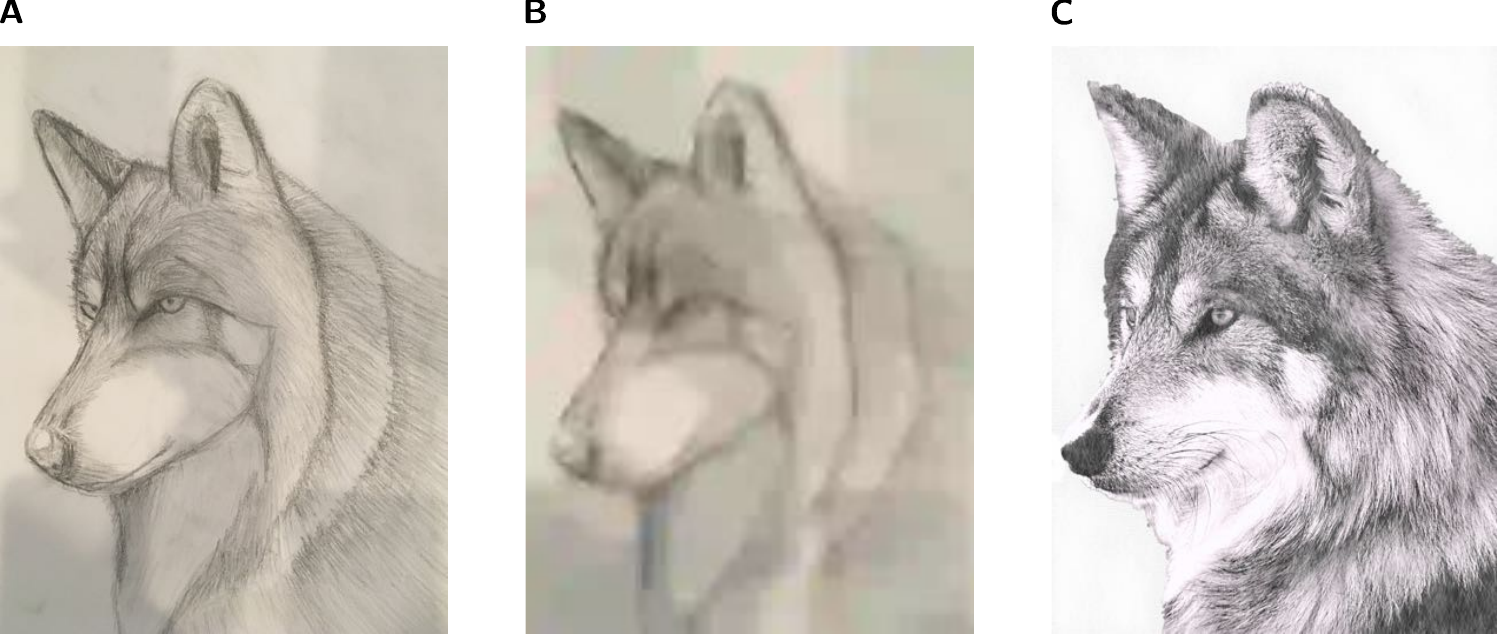}
\caption[Wolfsketch image results]{\label{figure:wolfsketch_results_appendix} (A) Original wolfsketch image with (B) WebP and (C) human reconstructions.}
\end{figure}

\end{document}